\documentclass[aps,amsfonts,amsmath,prd,preprint,nofootinbib,tightenlines,showpacs]{revtex4}
\usepackage{epsf}

\def\be{\begin{equation}}
\def\ee{\end{equation}}
\def\bea{\begin{eqnarray}}
\def\eea{\end{eqnarray}}
\def\bml{\begin{subequations}}
\def\blea{\bml\begin{eqnarray}}
\def\elea{\end{eqnarray}\end{subequations}}

\def\ba{\mathbf{a}}
\def\bb{\mathbf{b}}
\def\br{\mathbf{r}}
\def\bA{\mathbf{A}}

\begin{document}

\title{Cosmic string scaling in flat space}

\author{Vitaly Vanchurin}
\email{vitaly@cosmos.phy.tufts.edu}
\author{Ken Olum}
\email{kdo@cosmos.phy.tufts.edu}
\author{Alexander Vilenkin}
\email{vilenkin@cosmos.phy.tufts.edu}
\affiliation{Institute of
Cosmology, Department of Physics and Astronomy, Tufts University,
Medford, MA  02155}

\begin{abstract}
We investigate the evolution of infinite strings as a part of a
complete cosmic string network in flat space.  We perform a simulation
of the network which uses functional forms for the string position and
thus is exact to the limits of computer arithmetic.  Our results
confirm that the wiggles on the strings obey a scaling law described
by universal power spectrum.  The average distance between long
strings also scales accurately with the time.  These results suggest
that small-scale structure will also scale in expanding universe, even
in the absence of gravitational damping.
\end{abstract}


\maketitle

\section{Introduction }

During phase transitions in the early universe various topological
defects can form.  In particular, cosmic string networks are formed
when symmetry is broken and the vacuum manifold contains
incontractible loops \cite{Kibble}. Fundamental or $D$-strings formed
at the end of brane inflation can also play the role of cosmic strings
\cite{Tye,Polchinski,Gia}.  Primordial string networks can
produce a variety of observational effects: linear discontinuities in
the microwave background radiation, gravitational lensing, gamma ray
bursts, and gravitational radiation --- both bursts and a stochastic
background (for a review of cosmic strings, see \cite{VS,KH}).

An evolving string network consists of two components: long
strings and sub-horizon closed loops. The long string component is
characterized by the following parameters: the coherence length
$\xi(t)$, defined as the distance beyond which the directions along
the string are uncorrelated, the average distance between the strings
$d(t)$, and the characteristic wavelength of the smallest wiggles on
long strings, $l_{min}(t)$. The standard picture of cosmic string
evolution assumes that all three of these scales grow proportionately
to the horizon size $t$ and that the typical size of loops is set by
$l_{min}(t)$. 

Previous simulations of strings in an expanding universe
\cite{Bennett,Shellard} have only partially confirmed this model. The
long strings were observed to scale with 
\be 
\xi(t)\sim d(t) \propto t,
\label{standard}
\ee 
but the short wavelength cutoff $l_{min}$ and the loop sizes
did not scale, remaining at the resolution of the simulations. It is not
clear whether this is a genuine feature of string evolution or a
numerical artifact. A flat-space simulation introduced in \cite{Smith} and further developed in
\cite{Maria} and \cite{Hindmarsh} had the same problem. Moreover, the rate of growth of $\xi(t)$ and $d(t)$
in that simulation showed
dependence on the lower cutoff imposed on loop sizes, indicating that
lack of resolution at small scales can affect long string
properties.  

It is generally believed that in a realistic network the scaling
behavior of the cutoff scale $l_{min}(t)$ will eventually be enforced
by the gravitational back-reaction. However, it has been recently
realized that this back-reaction is much less efficient than
originally thought \cite{OS} and that the cutoff scale is sensitively
dependent on the spectrum of small-scale wiggles \cite{OSV}. Thus, it
becomes important to determine the form of the spectrum.

Here we have developed an algorithm and have performed a cosmic
string network simulation that lacks the problem of smallest
resolution scale.  Rather than representing the string as a series of
points which approximate its position, we use a functional
description of the position which can be maintained exactly (except
for the inevitable inaccuracy of computer arithmetic) in flat space.
After the simulation has run for some time and the total length of
string has decreased, we expand the simulation volume, as discussed
below.  This technique enables us to reach an effective box size
greater than 1000 times the initial correlation length.  Such a
technique could also be used in expanding-universe simulations.

The results of the simulation will be discussed in a series of
publications.  In the current paper we present the algorithm of our
simulation and concentrate on the evolution of infinite strings,
focusing in particular on the spectrum of small-scale wiggles.  In
section II we describe the algorithm of our simulation and its
complexity, in section III we show that the spectrum of small wiggles
exhibits scaling behavior.  In section IV we analyze the behavior of
the inter-string distance and correlation length, and show that they
scale with the time.  In section V we discuss what can be learned
about the expanding universe from this flat-space simulation.

\section{Numerical Simulation}

\subsection{Algorithm}

The evolution of a cosmic string with thickness much smaller than
radius of its curvature can be described by the Nambu action.  In
general, the equations of motion of cosmic strings cannot be solved
analytically.  However, in flat space-time they are greatly simplified
(see \cite{VS,KH} for detailed discussion).  The Nambu action is
invariant under arbitrary reparametrization of the two-dimensional
worldsheet swept by the string.  We take the time $t$ as one parameter
and a spacelike parameter $\sigma$ giving the position on the string.
With the usual parameter choice, the string trajectory
$\mathbf{x}(t,\sigma)$ obeys
\bea
\mathbf{\dot{x}}\cdot\mathbf{x}' & = & 0,\\
\dot{\mathbf{x}}^{2}+\mathbf{x}'^{2}&=&1,
\eea
with the equation of motion
\be
\ddot{\mathbf{x}}-\mathbf{x}''=0,
\ee
where prime denotes differentiation with respect to $\sigma$ and dot
differentiation with respect to $t$.

The general solution to these equations can be described by separating
left- and right-moving waves:
\be
\mathbf{x}(t,\sigma)=\frac{1}{2}\left(\ba(\sigma-t)+\bb(\sigma+t)\right),
\ee
with
\be\label{eqn:unitnorm}
\ba'^{2}=\bb'^{2}=1.
\ee

With this solution in hand we can follow the evolution of each string
in the network exactly.  Having computed $\ba$ and $\bb$
from the initial conditions we can find the position of each string
at any later time $t$.

When their trajectories intersect, strings can exchange partners with
some probability $p$.  For gauge theory strings, $p$ is essentially 1,
but fundamental strings and 1-dimensional D-branes can have
substantially lower $p$.  In the present analysis we are concerned
only with $p = 1$.  Smaller $p$ will be the subject of a future paper.

Such an intercommutation introduces discontinuities (kinks) in the
functional forms of the $\ba$ and $\bb$ describing the strings.  We
update the form accordingly and once again string positions can be
computed exactly.  This procedure was used by Scherrer and Press
\cite{SP} and Casper and Allen \cite{CA} to simulate the evolution
of a single loop; here we use it for a network.

In the expanding universe, small loops which are produced at
intercommutations will almost never encounter another string and so
will not rejoin to the network.  To capture this feature in a
flat-space simulation, we explicitly disallow loops of length
$l<\kappa t$ to rejoin the network, as was done in \cite{Maria, Hindmarsh}.
Here, $\kappa$ is a constant (typically taken as 0.25) and $t$ is the
time which has so far been simulated.  Of course the results of such a
technique can be trusted only if they are not sensitive to the
particular choice of $k$, and that expectation is confirmed below.

To generate initial conditions we use the Vachaspati-Vilenkin
prescription \cite{6} in a periodic box of size $L$, usually 100 in
units of the initial correlation length.  Where the string produced by
that technique crosses straight through a cell we use a straight
segment and where it enters a cell and exits through an adjacent face
we use a quarter circle.  However, to simplify the implementation we
replace the quarter circle with $K$ straight segments (connecting
points lying on the quarter circle).  In the present paper we use $K = 2$.
We expect our results to be insensitive to the choice of $K$, and in
fact tests with $K = 4$ showed no significant difference.

An initially static string of length $l$ will collapse into a double
loop at time $t = l/4$ \cite{KT}.  To prevent such a collapse we perturb the
initial conditions by giving small initial velocities to the part of
the string inside each cell in the normal direction to the local plane
of the string.

\subsection{Implementation}

To represent the piecewise linear form of $\ba$ and $\bb$ we store the
constant values of $\ba'$ and $\bb'$ for each segment, together with
values for $\ba(0)$ and $\bb(0)$.  To perform an intercommutation we
introduce 2 points for each of $\ba'$ and $\bb'$ and then concatenate
or split the lists of $\ba'$ and $\bb'$ depending on whether two
strings are joining into one or one string is splitting in two.

The world sheet of the piecewise linear string consists of flat pieces
that we call diamonds glued together along four lines each, two
constant values of $a(\sigma-t)$ and two constant values of $b(\sigma+t)$.
Each diamond has some extent in space and time.  At any moment, we
have a list of all diamonds that intersect the current time.  With
each diamond we store the spacetime position of the corners, so that
we don't have to integrate $\ba'$ and $\bb'$ to find the string
position.

When the current time exceeds the ending time of some diamond we
discard that diamond and generate a new one in the future light-cone
of the old one with the starting time of the new diamond equal to the ending
time of the discarded diamond.  When each new diamond is generated, we
check it for intersections with every existing diamond.  To accomplish
this in unit time, we divide the entire simulation box into small boxes
a little bit larger than the largest extent of any single diamond.
With each box we store a list of diamonds that intersect it.  Each
diamond can intersect at most 7 boxes.  When the diamond is created we
check each box that it intersects for possible intercommutations.

This algorithm does not necessarily generate intercommutations in
causal order, so when we detect one we store its parameters in a
time-ordered list of pending intercommutations.  When the current time
reaches the time of a pending intercommutation, the intercommutation is
then performed.  If instead we find an intercommutation in the
backward light cone of a pending one, we invalidate the pending one when
the earlier one is performed.

To be able to quickly determine the length of a newly created loop, we
store the $\ba'$ and $\bb'$ values in a ``skip list'' \cite{7}, which
permits integration of the displacements in $\sigma$ and updating for
intercommutations, both in time proportional to $\log l$.

The complexity of the above algorithm can be calculated in the
following matter.  In a box of size $L$ in units of initial
correlation length, $L^3$ random phases are generated.  There are
$3L^3$ edges and each of them has a probability of 8/27 to have a
string.  On average we have $(8/9)L^3$ of total string length in the
box of volume $L^3$.  Most of the string length is in a single long
string which is constantly intercommuting with itself and with other
loops.  To store $\ba'$ and $\bb'$ for all strings in the
network, we need on average $2(8/9)L^3K$ data points and about the
same number of diamonds.  Thus the memory usage scales as $L^3K$.

We use a ``calendar queue'' \cite{8} to store the list of diamonds,
which allows us to find the diamond with the earliest ending time and
to insert a new diamond into the list, both in constant CPU time.
Thus the entire procedure of finding the oldest diamond, removing it,
replacing it with a new diamond, and checking that diamond for
intercommutations requires only a constant amount of CPU time.  When
an intercommutation takes place, the runtime is proportional to the
logarithm of the length of the strings involved, because of the skip
list, but intercommutations are sufficiently rare that the time spent
searching for intercommutations dominates over the time spent
performing them.

The lifetime of a diamond depends on the lengths of the segments on
its edges, which are typically $1/K$.  Thus every time $1/K$ we need
to process of order $L^3K$ diamonds.  We typically simulate the
network for cosmic time of the order of the light crossing time of the box
$L$.  Therefore the final overall running time of the program
scales as $L^4K^2$.

Tufts has a Linux computation cluster on which we performed our
simulations.  Each of 32 nodes has a dual 2.8 GHz CPU and 3 GBytes of
usable memory.  Each run of the simulation runs on a single node.  The
memory constrains the box size to about 100 in initial correlation
length, and it takes about 5 hours to evolve the network for one light
crossing time of such a box.  The existence of multiple nodes allows
32 simulations to be run simultaneously, so that we can do about 150
runs in a day.

The code is written entirely in ANSI C, so it is easily portable
across different platforms.  It was successfully tested and ran on
Windows and Linux PCs.

\subsection{Expansion of the box}

In the previous section we have pointed out that the boundary effects
become important when the cosmic time of the simulation is of order
$L$. To avoid problems that have nothing to do with physics of the
process, we would like all of the length scales of the network to be
much smaller than the size of the box.  In the discussion below, we
propose a technique that allows us to overcome boundary effects and
push the effective box size and thus the running time of the
simulation to larger values.

Since small loops smaller than a threshold are not allowed to rejoin
the network, we can remove them from the simulation when they are
produced, and so decrease the the total string length.  The computer
memory used by simulation is proportional to the total string length
divided by the average size of the segments. Each intercommutation
introduces four kinks and thus decreases the average size of the
segments. This process increases the memory used by simulation by a
small constant amount. On the other hand when string loops decouple
from the simulation the memory is freed on average by the amount
proportional to their length. In the following discussion we shall
assume that the memory usage of the simulation is proportional to the
string length only, since we found that the decrease of the segment
size gives only a small correction to the calculations and does not
change the general idea.

Suppose that the string network is in the scaling
regime. If so, the inter-string distance grows linearly with $t$, and
thus the total string length in long strings decays as
$t^{-2}$. Therefore, by some time $t_1$, when the string length has
decreased by a factor of 8, the inter-string distance has grown by a
factor of $\sqrt{8}$. If we are only interested in behavior of
infinite strings we can expand the box at time $t_1$ from
initial size $L^3$ to $(2L)^3$ and continue the evolution further.

The simplest way to expand the box it to create 8 identical replicas
of the box of size $L^3$ at time $t_1$ and to glue them together to
form a box of size $(2L)^3$. This introduces correlations on the super
horizon distances that we should disturb by some mechanism to avoid
similar evolution of 8 boxes. If we do not do anything and continue
the evolution the result will not be different from evolution of a
single box where boundary effects become important at some time
of order $L$.

There are several ways that one can think of to disturb the
periodicity. The most promising approach that we have found, that does
least violence to the network is the following. Right after the
expansion we change the intercommutation probability to p=0.5 for some
fraction of the elapsed time and then change intercommutation
probability back to p=1. The time during which we force the
intercommutation probability to be a half should be at least the
inter-string distance in order to introduce enough entropy into the
system. The hope is that after $p=1$ is switched back on the system
will quickly return to the properties appropriate to $p = 1$

By the described procedure the expansion takes place roughly when the
total string length in long strings has decreased by 8. The average
time of expansions is the same regardless of the initial box size $L$,
so if we choose $L > t_1$ the first expansion would take place before
the boundary effects become important. However, the effective box size
grows in steps given by a power law $t^{2/3}$, and therefore
regardless of initial $L$ the effective box size would eventually
become smaller than current time. In order to be able to run the
simulation longer we have to be able to expand the box sooner. In fact
if we could have expanded the box when the total length in long
strings has decreased by 4, the effective box size would grow linearly
with time. This can be achieved if we also modify the string at the
time of expansion by removing half of the points of long strings to
increase the average size of the segments by two.

Each string consists of about the same number of left and right moving
kinks, and our job is to replace it with half as many points that
represent a smoothed version of the original string. One possibility
is to remove three out of every four kinks (both left and right
moving) and connect the remaining points with straight segments. Then
each of the remaining points is decomposed into two kinks, one left
and one right moving.  At this point we still have a freedom to give
new segments any velocities we like in the direction orthogonal to the
tangent vector of the string. To set each new segments with new
velocities we find an average velocity of old four segments and
project it to the transverse plane of the new segment. This algorithm
shortens the string length since we replace four consecutive segments
with a single straight segment. Although the inter-string distance
defined as a $1/\sqrt\rho$ becomes larger after smoothing, the real
inter-string distance, that describes the average distance between
nearby long strings, does not change.

The expansion of the box technique with smoothing can be carried on
over and over again until we run into another problem of periodic
boundary conditions, but the effective box size would be somewhat
larger. The new problem has to do with the fact that overall linkage
of the initial box is zero. However, we would like to think of our box
as a small part of an infinite space. This means that the overall
linkage of the box should be some random number with distribution
peaked around zero, but not identically zero. On the other hand, the
zero linkage of the box results in very odd configurations of the network
at late times: for any long string with linkage in, say, the positive
$x$ direction, there must be another string close by linked in the
negative $x$ direction. This effect leads to overproduction of small
loops, because the configuration is not stable and long strings would
tend to unlink themselves.  It prevents us from expanding the box
indefinitely.

\section{Scaling Properties}

We will primarily be concerned with the amplitude of small wiggles
on infinite strings.  To have a time-independent description of these
wiggles, we will consider wiggles in the functions $\ba$ and
$\bb$.  With a very long loop of string, we can define a
Fourier transform of the tangent vector,
\be\label{eqn:Fourier}
\bA(k)=\frac{1}{\sqrt{2\pi l}}\int_{0}^{l}d\sigma\, 
\ba'(\sigma)e^{-ik\sigma}
\ee
where $l$ is the loop length, and $k = k_n = 2\pi n/l$ for $n$ an integer.
We will work in the limit where $l\to\infty$, so that $k$ becomes a
continuous variable, but because our functions do not fall off at
large $\sigma$, we will have to retain the parameter $l$.

There will be some level of correlation between different modes.  For
example, since modes of similar size can collectively form a loop we
would expect a large mode which had survived on a long string to be
accompanied by smaller than average modes of similar wavenumber.
Nevertheless, we can get some idea of the string shape by ignoring
such correlations and taking
\be\label{eqn:2point}
\langle \bA(k)\cdot \bA^*(k')\rangle = \begin{cases}c(k) & k = k'\\
0 & k\neq k'\end{cases}
\ee
where the averaging is over realizations of the string network.
Because of the factor $1/\sqrt{l}$ in  Eq.\ (\ref{eqn:Fourier}),
$c(k)$ does not depend on $l$,  
\be
c(k)=\frac{1}{2\pi}\int_0^l d\sigma e^{ik\sigma}C(\sigma),
\label{c}
\ee
where $C(\sigma)$ is the correlation function
\be
C(\sigma)=\langle \ba'(0)\ba'(\sigma)\rangle.
\label{csigma}
\ee
The values $\bA(k)$ for different $k$ have no correlation.  In the
limit $l\to\infty$, these different points become arbitrary close, so
$\bA(k)$ is everywhere discontinuous.  However, $c(k)$ is a continuous
function, and $c(-k)= c(k)$ since $C(\sigma)$  is real.

The inverse transformation to (\ref{eqn:Fourier}) is
\be
\ba'(\sigma) =\sqrt{\frac{2\pi}{l}}\sum_{n =-\infty}^\infty
e^{ik_n\sigma}\bA(k_n) 
\ee
From  Eqs.\ (\ref{eqn:unitnorm}) and (\ref{eqn:2point}),
\be
1 =\langle\ba'(\sigma)^2\rangle
= \frac{2\pi}{l}\sum_{n =-\infty}^\infty c(k_n)
\approx\int_{-\infty}^\infty dk\, c(k)
\ee
Similar relations apply for $\bb'(\sigma)$.

We can define a ``spectral power density'',
\be\label{eqn:P}
P(k) = 2 k c(k)
\ee
so that
\be\label{eqn:Pint}
\int_0^\infty P(k)\, d\ln k = 1\,.
\ee
The quantity $P(k)$ is the fractional contribution to the energy density of
the string from modes in a logarithmic interval around $k$.

\begin{figure}
\begin{center}
\leavevmode\epsfxsize=4in\epsfbox{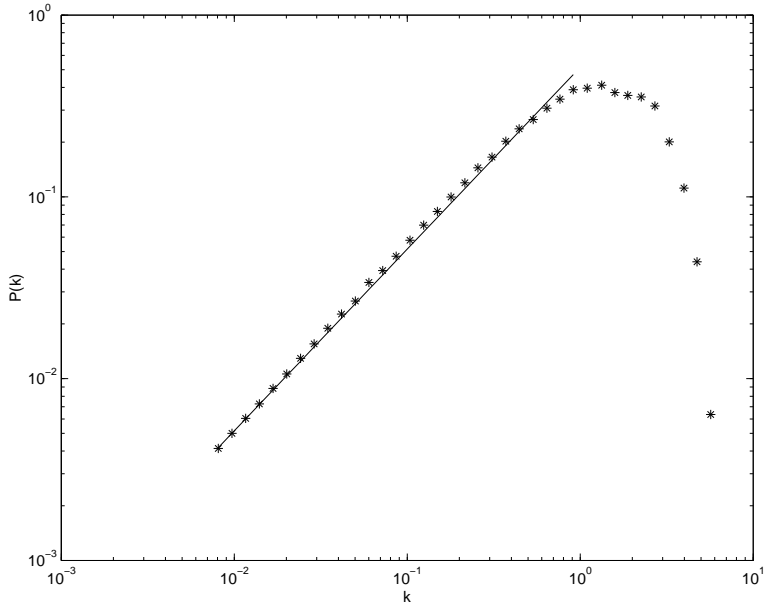}
\end{center}
\caption{Initial power spectrum.  The spectrum at large scales is well
fit by the random walk spectrum with step size $h= 1.62$.}
\label{fig:initial}
\end{figure}

For Vachaspati-Vilenkin initial conditions it has been shown that
properties of long strings are similar to properties of a random walk
\cite{6}.  Computing the power spectrum for a random walk of step size
$h$ gives
\be
P(k) =\frac{4}{\pi kh}\sin^2 (kh/2)\to
\begin{cases}kh/2& kh\ll 1\\
2/(\pi kh)\text{ on average}& kh\ll 1\end{cases}
\ee
We create our initial conditions with straight segments of length 1
and piecewise linear approximations to curved segments of length
$\pi/4$.  However, we have some correlation between successive
segments, so the effective correlation length of the random walk in
the initial conditions is $h\approx 1.62$ as shown in Fig.\
\ref{fig:initial}.
We do not attempt here to fit the small-scale
power spectrum, which depends on the detailed choice of the piecewise
linear string.

How should the structures on a string evolve at late times?  If there
is a scaling regime, and if a certain
fraction of the power is between wave numbers $k$ and $k'$ at time $t$,
then the same fraction of the power will be found between
wavenumbers $k/\lambda$ and $k'/\lambda$ at time $\lambda t$.  This
implies that $P(k)$ is a function of $kt$ alone.

Any function $P(kt)$ would give scaling, but we can make some guesses
about the shape of this function.  First of all, on scales larger than
horizon (i.e., $kt\gg1$), we have a random walk, so $P(kt) \propto
kt$.  The function $P(k)$ outside the horizon is increasing with time,
but that is just a result of the smoothing of the string on smaller
scales and our definition of the power.

Inside the horizon, we expect loop production to smooth the string.
If there are large excursions in $\ba$ and $\bb$ at any scale, we
expect them eventually to overlap with each other in such a way as to
intercommute and produce a loop.  The loop will then be removed from
the string and the power at that scale will decrease because it has
gone into the loop.  The typical angle that such excursions make with
the general direction of the string is given by an integral of $P(k)$
over a range of wavenumbers describing the size of the wiggle.  We
thus expect that this process will reduce $P(k)$ below some critical
value independent of $k$, at which the wiggles are large enough to
produce loops.

If that were the only smoothing process, we would expect a constant
spectrum $P(kt)$.  But even very small-amplitude structure can be
smoothed by other processes, such as loop emission at cusps.  If the
large-scale structure would produce a cusp, then small wiggles can
produce instead a self-intersection that emits a loop.  The process is
similar to the emission of vortons at cusps \cite{vortongun}.  If a
region of string has particularly large wiggles, it is more likely to
be emitted into a loop by this process, so there will be gradual
smoothing even on scales that are already quite smooth.  Thus we might
expect some decline below the flat spectrum.

\begin{figure}
\begin{center}
\leavevmode\epsfxsize=5in\epsfbox{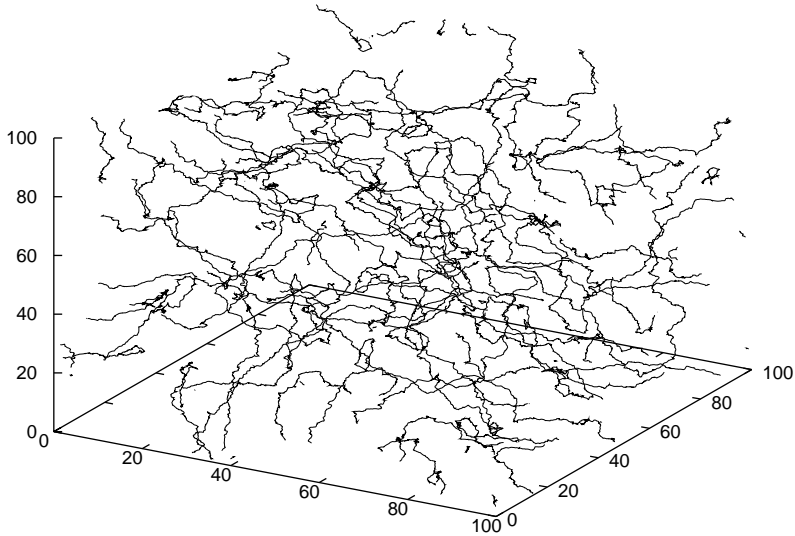}
\end{center}
\caption{The network of long strings at time 100 of a simulation run in
a box of size 100 in units of initial correlation length.}
\label{fig:strings}
\end{figure}

\begin{figure}
\begin{center}
\leavevmode\epsfxsize=4in\epsfbox{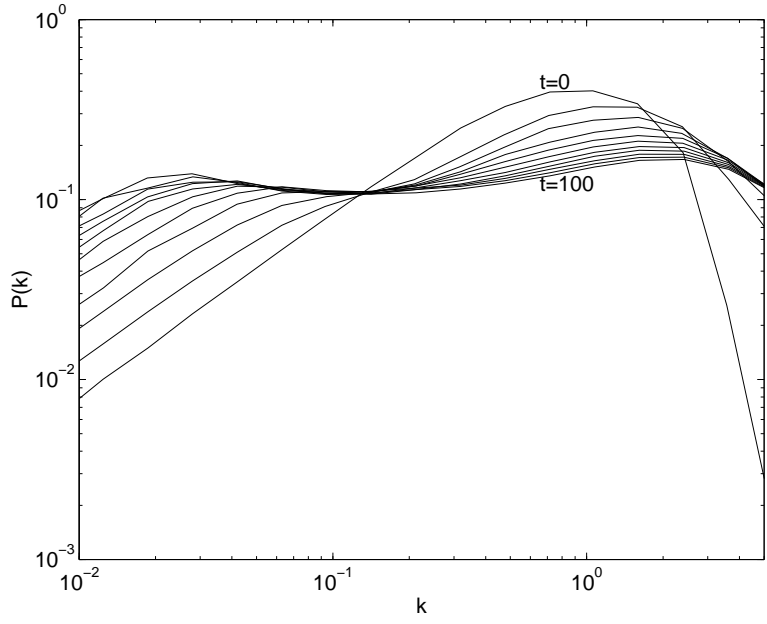}
\end{center}
\caption{Evolution of the power spectrum from the initial conditions
 in steps of 10 to $t = 100$  in a box of size 100. }
\label{fig:spectra}
\end{figure}

In Fig.\
\ref{fig:strings},
we show a picture of the network of long strings at the end of a run without expansion of the box technique,
and in Fig.\
\ref{fig:spectra},
we see the power spectrum at different times.  The production of small
loops constantly reduces the power on the scale of the initial
correlation length.

\begin{figure}
\begin{center}
\leavevmode\epsfxsize=4in\epsfbox{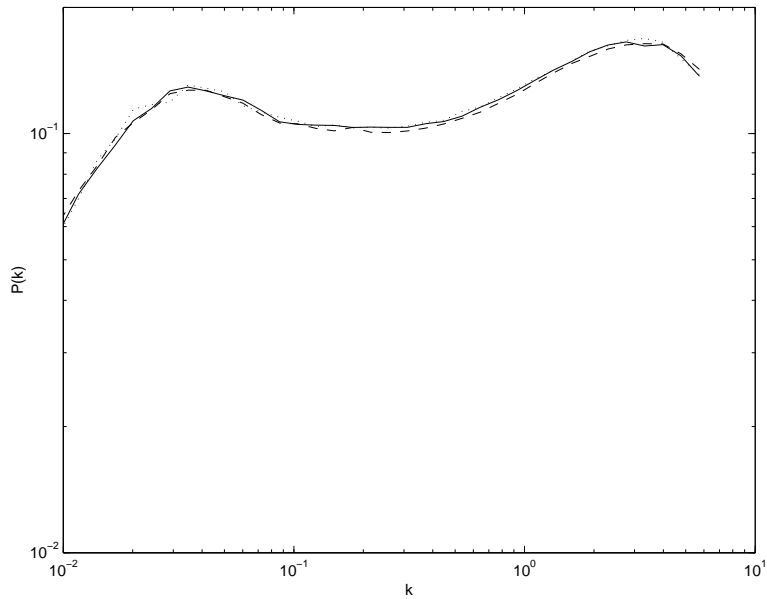}
\end{center}
\caption{Final power spectrum for rejoin cutoff constant $\kappa = 0.125$
(dashed), 0.25 (solid), and 0.5 (dotted).}
\label{fig:rejoin}
\end{figure}
In Fig.\
\ref{fig:rejoin}, we compare the power spectra at $t =
100$ for various values of the cutoff $\kappa$ which controls how small a
loop is allowed to rejoin the network.  We see that the choice of this
cutoff has little effect on the results.  In fact, even allowing all
rejoining has a fairly minor effect.

\begin{figure}
\begin{center}
\leavevmode\epsfxsize=4.5in\epsfbox{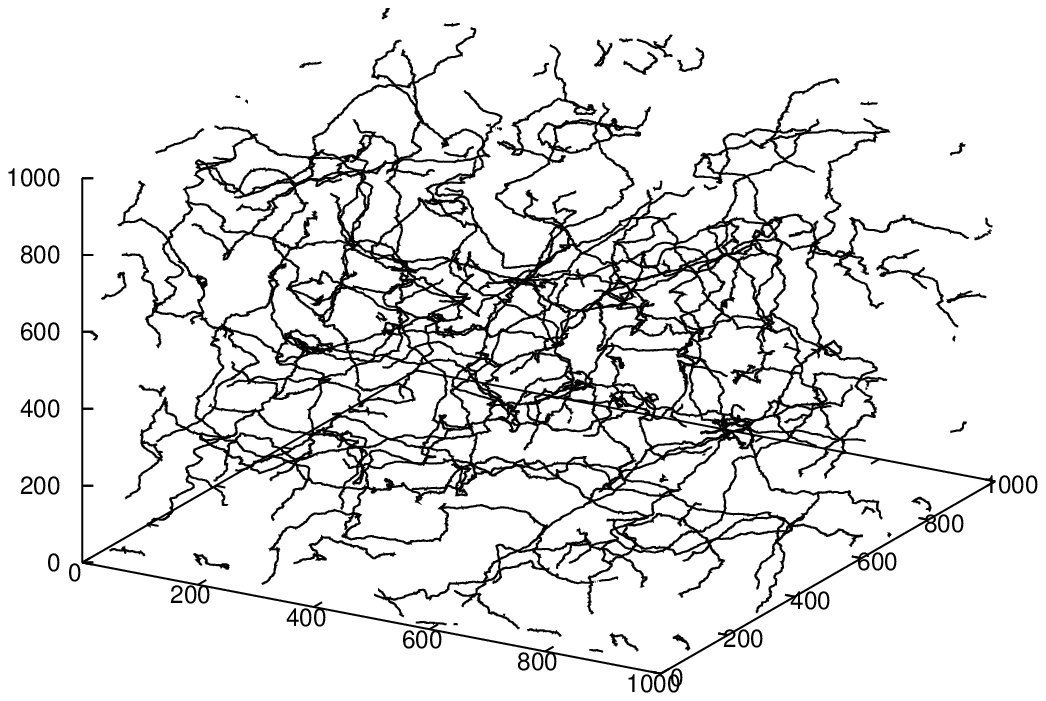}
\end{center}
\caption{Part of the network of long strings at time 960 after 4
doublings of the box size.}
\label{fig:new-strings}
\end{figure}

\begin{figure}
\begin{center}
\leavevmode\epsfxsize=4.5in\epsfbox{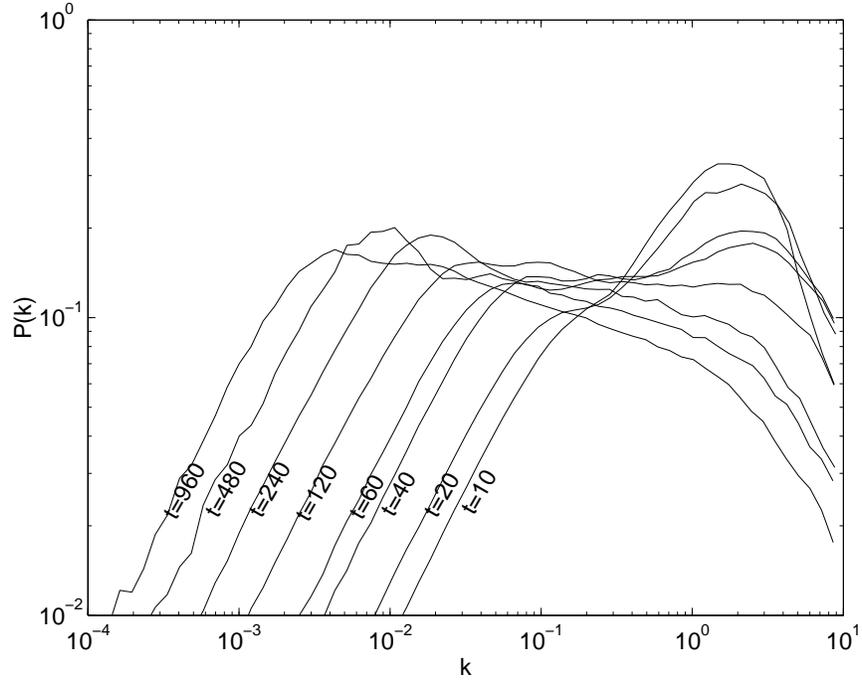}
\end{center}
\caption{Evolution of the power spectrum vs.\ $k$ at late times with expansion of the box technique}
\label{fig:new-spectrum-k}
\end{figure}

\begin{figure}
\begin{center}
\leavevmode\epsfxsize=4.5in\epsfbox{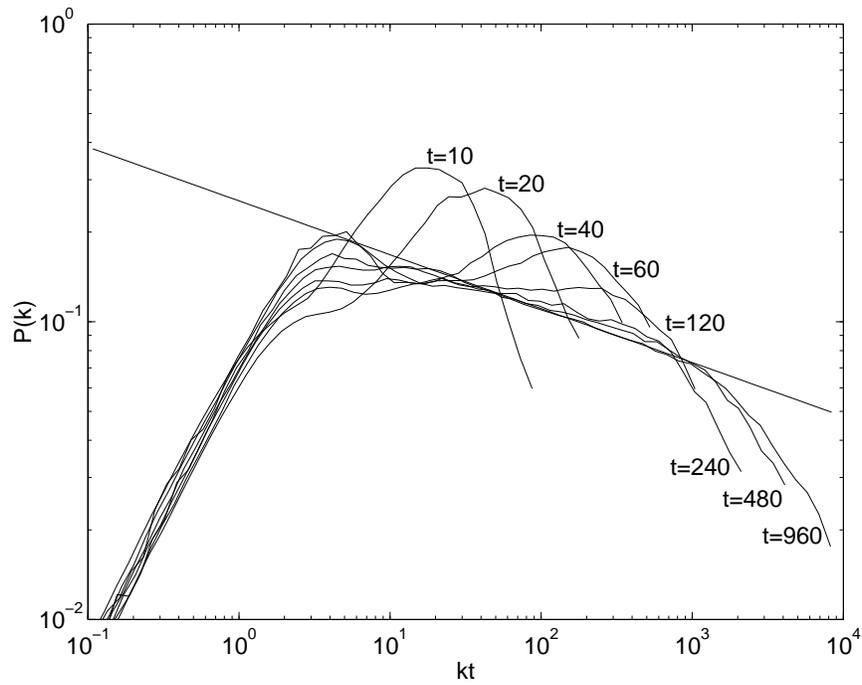}
\end{center}
\caption{Evolution of the power spectrum vs.\ $kt$ at late times with expansion of the box technique}
\label{fig:new-spectrum-kt}
\end{figure}

In Fig.\
\ref{fig:new-strings},
we see a picture of the network of long strings at time 960, after 4
doublings of the box size,
and in Fig.\
\ref{fig:new-spectrum-k},
we plot the evolution of the power spectrum. We also show the
evolution of power spectrum vs.\ $kt$ in Fig.\ \ref{fig:new-spectrum-kt},
to demonstrate that all spectral lines cluster along some universal
power spectrum. This demonstrates that the power spectrum at late
times is a function of $kt$ alone, just like we would expect for a
scaling network. The slowly declining part of the spectrum has a small
power law dependence, which is presumably a result of processes such
as loop emission near cusps.  The best fit power law is
\be\label{eqn:powerfit}
P(kt)=0.25 (kt)^{-0.18}\,,
\ee
only a small correction to the flat spectrum.

We cannot see this power law extend to arbitrarily large $kt$, since
we are smoothing the string at small scales as part of the expansion
procedure.  But we conjecture that without this smoothing we would
find the form shown in Fig.\ \ref{fig:new-spectrum-kt} with the power
law decline 
eventually extending to arbitrarily large $kt$. In that case, at late times the power spectrum
will fall off smoothly to very low levels, rather then coming to a
sudden end at some scale $l_{min}$.  Because
the exponent in Eq.\ (\ref{eqn:powerfit}) is negative there will be no small-scale divergence in
the integral in Eq.\ (\ref{eqn:Pint}).

\section{Length Scales}

From the discussion above we expect structures on each string to
scale, and we also expect that the properties of the network as a
whole should exhibit scaling. Thus any characteristic length must be
proportional to the time $t$.

\begin{figure}
\begin{center}
\leavevmode\epsfxsize=4.5in\epsfbox{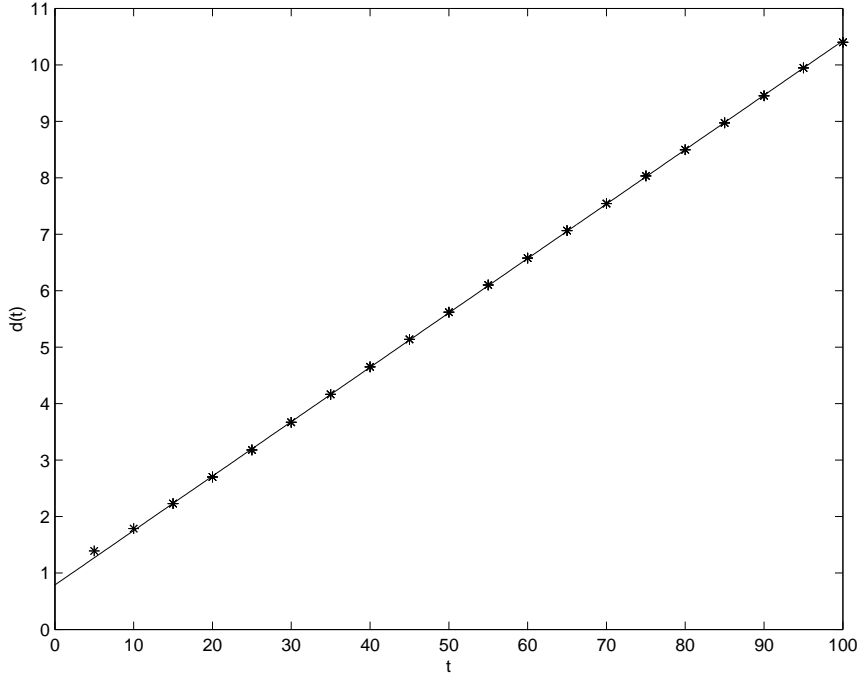}
\end{center}
\caption{Inter-string distance $d(t)$ at various times (points).  The
line is $d(t) = 0.096(t+8.2)$.}
\label{fig:interstring}
\end{figure}

We will consider two such lengths.  The first is the average distance
between strings, which we define in a simple fashion \cite{ACK}
\be
d(t)=\sqrt{1/\rho}\,.
\ee
where $\rho$ is the length of infinite strings per unit volume.
In the case of our simulation box of size $L$,
\be
d(t)=\sqrt{L^3/{\cal L}(t)},
\ee
where ${\cal L}(t)$ is the total string length of long strings.
To compute $d(t)$, we have to specify which strings are ``long'',
but fortunately the result is insensitive to the exact definition.
The result for length cutoff $0.25t$ is shown in Fig.\
\ref{fig:interstring}.
The distance is a linear function of $t$, but the $t$-intercept is not
at $t = 0$.  Instead the best fit line is
\be
d(t)=0.096(t+8.2)\,.
\ee
Thus it appears that, at late times, the typical string distance is only
about 1/10 of the elapsed time since the start of the simulation (here
called $t = 0$).  Since our initial conditions had inter-string distance
about 1, they correspond to an initial setting of the natural time
parameter about 10.

\begin{figure}
\begin{center}
\leavevmode\epsfxsize=4.5in\epsfbox{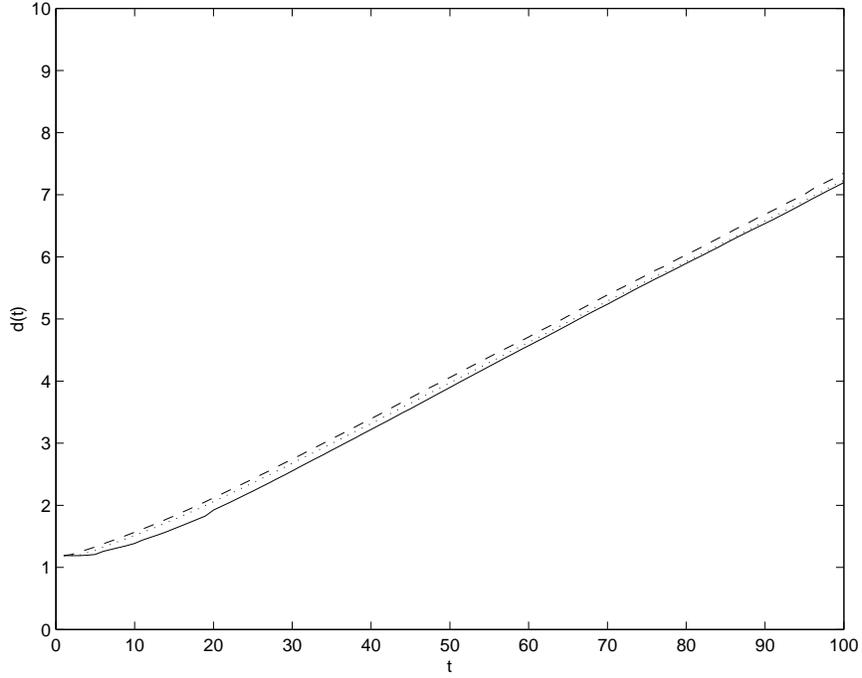}
\end{center}
\caption{Average distance between long strings $d(t)$ computed by
considering long strings to be those longer than $0.25t$ (solid),
$1.0t$ (dotted) and $5.0$ (dashed).}
\label{fig:longcutoff}
\end{figure}

In Fig.\ 
\ref{fig:longcutoff},
we show the inter-string distance
computed for various values of the threshold used to define
``long'' strings.  We see that changing this parameter has little
effect on the distance computation.

\begin{figure}
\begin{center}
\leavevmode\epsfxsize=4.5in\epsfbox{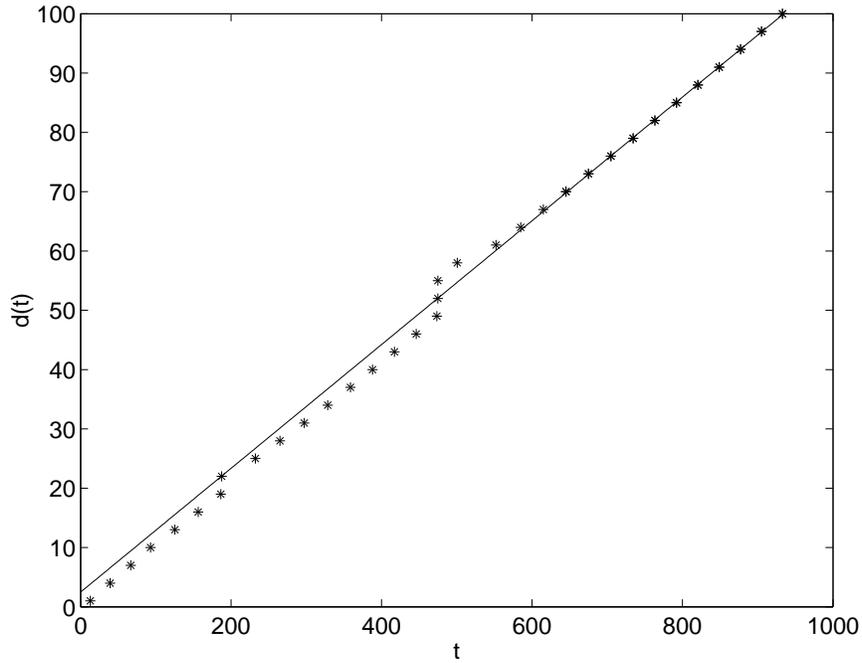}
\end{center}
\caption{Inter-string distance $d(t)$ with expansion of the box.}
\label{fig:new-interstring}
\end{figure}

With expansion of the box we can run for much longer. Fig.\ 
\ref{fig:new-interstring}
shows the inter-string distance out to time 1000.  The evolution
remains linear, but there is a jump each time the box is expanded.

The second scale is a correlation length $\xi(t)$, such that the
tangent vectors to the string at two points separated by distance
$\xi(t)$ still have significant correlation.  To minimize the
dependence of this measure on structure at very small scales, we
proceed as follows.  Let $\br(\sigma, l)$ be the integral of the tangent
vector $\ba'$ between $\sigma$ and $\sigma +l$, or equivalently the
total displacement in $\ba$ between those points,
\be
\br(\sigma, l) = \ba(\sigma)-\ba(\sigma+l)
\ee
One can use $\bb$ in precisely the same way.  A measure of the
correlation at separation $l$ is given by the dot product between
successive segments,
\be
C(l) =\langle\hat\br(\sigma, l)\cdot\hat\br(\sigma+l, l)\rangle
\ee
where $\hat\br=\br/|\br|$.  The average physical separation between
points on scales $l$ is given by
\be
r(l) =\langle|\br(\sigma, l)|\rangle\,.
\ee
We define the correlation length $\xi(t)$ to be $r(l)$ at the scale
where $C(l) = 0.2$.  The choice of the relatively small value 0.2 is
motivated by attempting to reduce the contribution of small scales as
much as possible while still having a threshold that one can clearly
distinguish from no correlation.

\begin{figure}
\begin{center}
\leavevmode\epsfxsize=4.5in\epsfbox{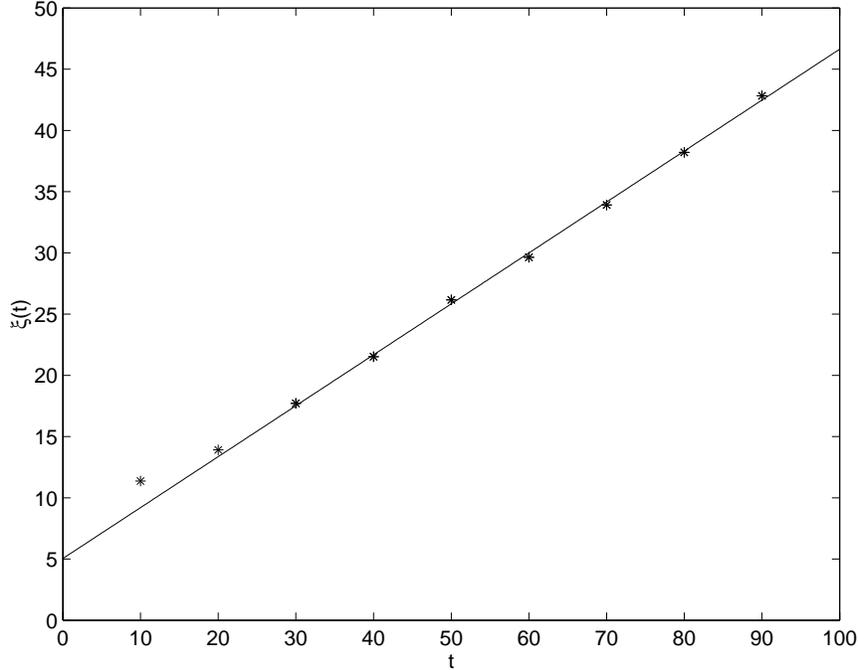}
\end{center}
\caption{Correlation length $\xi(t)$.  The evolution is very nearly
linear with $\xi(t)\sim 0.41t$.}
\label{fig:correlation}
\end{figure}

\begin{figure}
\begin{center}
\leavevmode\epsfxsize=4.5in\epsfbox{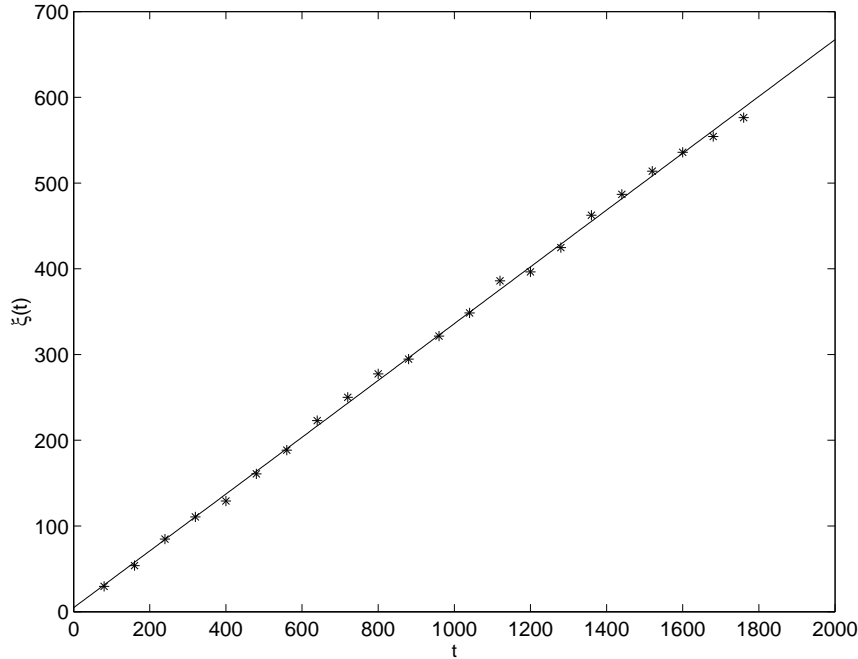}
\end{center}
\caption{Correlation length $\xi(t)$ with expansion of the box technique. The best-fit straight line is $\xi(t) \sim 0.34t$}
\label{fig:new-correlation}
\end{figure}

In Fig.\ 
\ref{fig:correlation} we show the linear growth of $\xi(t)$. The best-fit straight line is
\be
\xi(t)\sim 0.41t\,.
\ee
The same quantity is plotted in Fig.\ \ref{fig:new-correlation} with
box expansion.

A different correlation length scale was defined by Austin, Copeland,
and Kibble \cite{ACK}, who used
\be
\bar\xi=\int_0^\infty d\sigma\, C(\sigma)
\ee
The scale behaves similarly to $\xi(t)$ in our simulation, except
that there are significant jumps at each expansion because smoothing
eliminates the small-scale structure.

One way to interpret this situation is to take the part of the run before the
final expansion as preparation of initial conditions for the part that
comes afterward.  Our initial configuration was not very close to the
configuration of the scaling network because there is too much
structure at the initial correlation length.  After the last expansion
if our choice of the initial condition matches the one of the scaling
network we would get into the scaling regime fast enough before
running into the problem of finite box size. With this interpretation
we would not have to worry too much about jumps in $d(t)$ (or in $\xi(t)$, although the definition above mostly eliminates those).

\section{Discussion}

Given our results for flat space, what can we say about the situation
in an expanding universe?  We might hope that our results could be
taken over into an expanding space-time with our spatial and temporal
coordinates becoming conformal distances and comoving time, but there
is potentially an important difference because the expansion of the
universe can damp the oscillating wiggles.

A simple model \cite{OSV} would be to imagine that intercommutations
act at the horizon scale to produce a spectrum in which $P$ depends
only on $kt$ as we found above.  A mode thus enters the horizon with
$P(kt) = P$ when $k\sim 1/t$, where $t$ is the conformal time.
Assuming that the mode is not affected by larger-wavelength modes that are being
damped as in \cite{9}, its comoving $k$ is unchanged. Comoving mode
amplitudes, however, decrease with the increase of the scale factor,
so we find the amplitude goes as
\be
\frac{a_k}{a_0} = \left(\frac{t_k}{t_0}\right)^\alpha \sim
\frac{1}{(kt_0)^\alpha}
\ee
where $t_k$ is the time where the mode entered the horizon, $a_k$ is
the scale factor at that time, $t_0$ and $a_0$ are the present time
and scale factor, and the exponent $\alpha$ gives the dependence of
scale factor on time, $a(t) \sim t^\alpha$ with $\alpha = 1$ in the
radiation-dominated universe and $\alpha = 2$ in the matter-dominated
universe.  The amplitude appears squared in $P(k)$, so in this model
we would find $P(k)\sim k^{-2}$ for radiation dominated and $k^{-4}$
for matter dominated.

This model is overly simplistic, and in reality the interactions
between intercommutations and expansion are more complex.
Nevertheless we expect that expansion can only make the string
smoother than in the flat-space case, so there will be a scaling
solution in the expanding universe, even without gravitational
damping.  We expect the spectrum to have a power law form with
exponent between the values above and the flat space result given by
Eq.\ (\ref{eqn:powerfit}).

In upcoming publications we will discuss the questions of loop
production and fragmentation, the average velocity and effective mass
density of long strings, and the evolution of strings with different
intercommutation probabilities $p<1$.

\section*{Acknowledgments}

We are grateful to Noah Graham for helpful discussions and to
Jordan Ecker for help with part of the computational system.  This
work was supported in part by the National Science Foundation.

\end{document}